\documentclass[aps,pra,twocolumn,superscriptaddress]{revtex4-1}

\usepackage[latin9]{inputenc}
\usepackage{color}
\usepackage[colorlinks,urlcolor=blue,citecolor=blue,linkcolor=blue]{hyperref}
\usepackage{verbatim}
\usepackage{float}
\usepackage{amsmath}
\usepackage{amssymb}
\usepackage{graphicx}
\usepackage{epstopdf}
\usepackage{esint}
\usepackage{CJK}
\usepackage{braket}
\begin{document}
\begin{CJK*}{UTF8}{gbsn} 
\title{BEC immersed in a Fermi sea: Theory of static and dynamic behavior\\ across phase separation}



\author{Bo Huang (黄博)}
\affiliation{Institut f\"ur Quantenoptik und Quanteninformation (IQOQI), \"Osterreichische Akademie der Wissenschaften, 6020 Innsbruck, Austria}
\affiliation{Institut f\"ur Experimentalphysik und Zentrum f\"ur Quantenphysik, Universit\"at Innsbruck, 6020 Innsbruck, Austria}

\date{\today}
\begin{abstract}
We theoretically study the static and dynamic behavior of a BEC immersed in a large Fermi sea of ultracold atoms under conditions of tunable interspecies interaction. The degenerate Bose-Fermi mixture is kept in an elongated trap, typical for a single-beam optical dipole trap.
We focus on the case of repulsive Bose-Fermi interaction and develop mean-field models to simulate the system over a wide range of repulsion strength. We further get analytical solutions in the regimes of phase separation and weak interaction. We obtain static density profiles and the frequency of the radial breathing mode, which is an elementary dynamic phenomenon of the mixture. Our results unveil the structure of the Bose-Fermi interface and describe the origin of the frequency shift of the breathing mode when the components become phase-separated at strong repulsion. We show that the mediated interaction between bosons induced by the Fermi sea can be understood as an adiabatic second-order mean-field effect, which is valid also beyond the weak-interaction regime for relevant experimental conditions. These results are consistent with our recent observations in a mixture of $^{41}$K and $^6$Li. 
\end{abstract}


\maketitle
\end{CJK*}



\section{Introduction}
Historically, studies on multi-component quantum fluids were conducted on mixtures of $^3$He with $^4$He \cite{Ebner1971tlt} and hydrogen with deuterium or tritium~\cite{Stwalley1976pnq}. The more recent achievements on ultracold atomic gases~\cite{Pethick2002book,Pitaevskii2016bec}, in which the interatomic interactions can be tuned by Feshbach resonances~\cite{Chin2010fri}, offer experimentalists many opportunities to create and study interacting quantum mixtures of different spin states, isotopes, or species. One situation is mixing  quantum fluids of different quantum statistics: degenerate Fermi gases (DFG) and Bose-Einstein condensates (BEC)~\cite{Truscott2001oof,Schreck2001qbe}. In these mixtures, interspecies interactions between fermions and superfluid BECs lead to rich phase diagrams (e.g.~\cite{Ludwig2011qpt}) and dynamic phenomena~\cite{Pitaevskii2016book}.

The equilibrium state of degenerate Bose-Fermi mixtures and their stability have been studied in experiments~\cite{Modugno2002coa,Ospelkaus2006idd,Lous2018pti} and also in mean-field theories, in which often the zero-temperature case is considered~\cite{Suzuki2008psa,
Marchetti2008psa}. It is a well-established fact that the constituents of a mixture undergo phase separation when the interaction is repulsive and strong, and the BEC collapses when the interaction is substantially attractive. An important phenomenon of phase separation is the formation of a thin interface between the components, and the remaining boson-fermion overlap at the interface has recently been probed by measuring three-body recombination losses in a $^{41}$K-$^6$Li mixture~\cite{Lous2018pti}. Therefore we are encouraged to investigate the detailed structure and properties of the thin interface theoretically.

Generally, collective excitations of the BEC can be described by the Gross-Pitaevskii equation (GPE)~\cite{Pitaevskii2016bec}, and the evolution of the DPG is analyzed either with kinetic equations in a semiclassical manner or via response functions of perturbations~\cite{Brack1997sp,Fetter2003qto}. The latter is rather different from superfluid fermions~\cite{Pitaevskii2016book}, which can be described by order parameter and hydrodynamic equations. 
In a miscible mixture, inter-species interaction is modeled as meanfield to study elementary excitations of zero-temperature Bose-Fermi mixtures in homogeneous~\cite{Yip2001cmi} and harmonically trapped~\cite{Vichi1998qda} conditions. In the limit of full phase-separation, BEC and DFG establish pressure balance at the interface, and their collective modes in a harmonic trap have been investigated~\cite{Lazarides2008ceo,Schaeybroeck2009tps}.
There also have been efforts to understand the dynamics over a range of interaction strengths, and Ref.~\cite{Maruyama2005moa} constructed a numerical model for the time evolution of monopole oscillations in a Bose-Fermi mixture. 

Experiments on the center-of-mass (COM) mode of the BEC coupled to the fermions show very weak frequency shifts on the order of a few percent~\cite{Ferrier2014amo,Delehaye2015cva,Wu2018cdo,Roy2017tem,DeSalvo2019oof}. However, our recent measurements on the radial breathing mode (RBM) frequency demonstrate a striking frequency shift of up to about 40 percent at phase separation~\cite{Lous2018pti}, whereas the frequency remain almost unchanged when bosons and fermions are mixed.

In this work, we consider the situation in our recent experiments ~\cite{Lous2018pti,Huang2019bmo}. A small BEC is immersed in a large Fermi sea and both components are trapped in an elongated harmonic trap. We first set up a mean-field model and numerically calculate the static density profiles of the Bose-Fermi mixture at zero temperature and extract the overlap between the two components at various strength of Bose-Fermi repulsion. Then we obtain analytical descriptions of the density profiles in the regimes of weak interaction and phase separation, and hence get insight into the formation and structure of the Bose-Fermi interface. We also discuss the weak finite-temperature effects on the Bose-Fermi overlap.

As an elementary example of dynamic behavior, RBM of the BEC immersed in the DFG is theoretically studied in the current work. We consider the condition of our experiments \cite{Huang2019bmo} on RBM, and the dynamics of DFG is  investigated in three different ways. 
In the first case, we assume that the DFG adiabatically follows the oscillation of the BEC. Our second approach considers single fermion trajectories in the full phase-separation limit. Finally, we also perform numerical simulations for fermions using the test-particle method. Our results show that, when the Bose-Fermi $s$-wave scattering length $a_{bf}$ increases from zero, the RBM frequency $\omega$ remains almost constant until the fermions are depleted from the trap center by the BEC. Then the frequency increases dramatically across the phase separation until it levels off at the full phase-separation limit. The relation between the plateau value of $\omega$ and the number ratio $N_b/N_f$ unveils the essential role of the compressional character of the breathing mode.

In Sec.~\ref{sec:static}, we first present our mean-field model, which has been used in Refs.~\cite{Lous2018pti,Huang2019bmo}, and  study the static properties of a Bose-Fermi mixture with tunable repulsive interspecies interaction. Then we obtain analytical results for the situations of weak interaction and phase separation, and, in particular, intuitively explain the structure of the interface in immiscible Bose-Fermi  mixtures. In Sec.~\ref{sec:dynamic}, we develop three models to study the radial breathing mode of a BEC immersed in a large Fermi sea, and the frequency of RBM at different strengths of repulsion is calculated. We compare our results from different models and discuss the validity of the adiabatic-Fermi-sea approximation. In Sec.~\ref{sec:conclusions}, we summarize our results and their impacts.

\section{Static properties}\label{sec:static}
In our recent work  \cite{Lous2018pti,Huang2019bmo}, we developed a numerical mean-field model to calculate static density profiles of Bose-Fermi mixtures at zero-temperature. At finite temperatures, we use the density of BEC and fermions to calculate mean-field potential and estimate the thermal-bosons density. Here we first describe in detail this model, of which the basic ideas have been mentioned in the Supplemental Material of Ref.~\cite{Lous2018pti}. Then we calculate the Bose-Fermi overlap $\Omega$, which is an experimental observable. We further obtain analytical results for $\Omega$ at weak interaction and full phase separation. Finally, finite-temperature effects on densities are estimated.

\subsection{Numerical model}
\label{sec:static_numeric}
\begin{figure*}[ht]
\centering
\includegraphics[scale=0.8]
{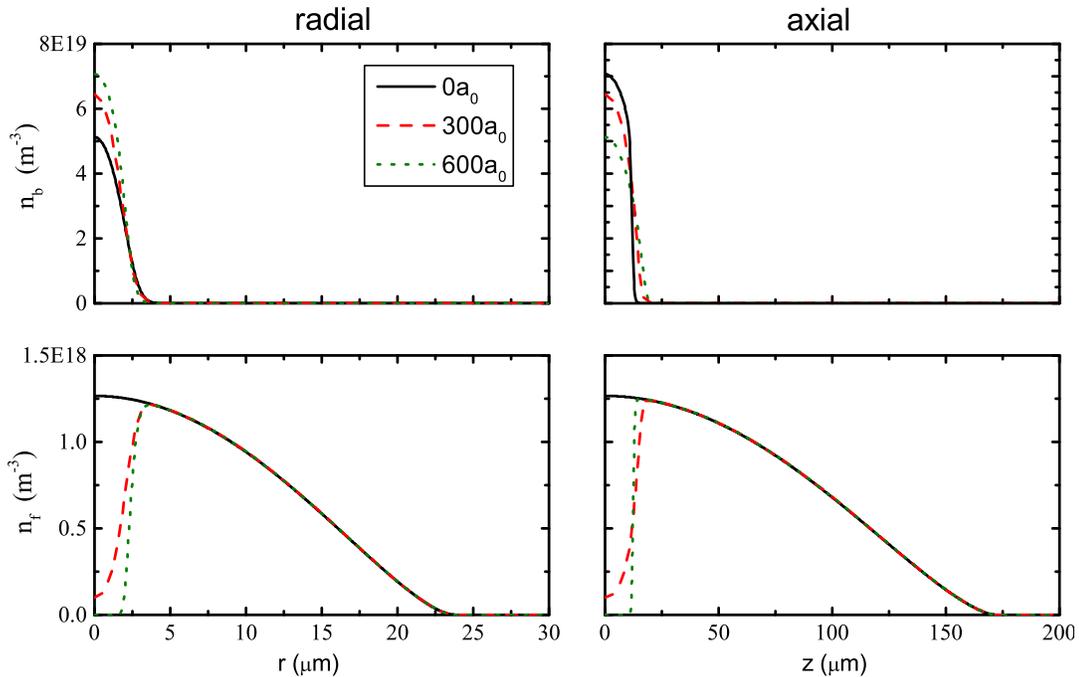} 
\caption{The density profiles of the BEC ($n_b$) and the fermions ($n_f$) along the radial ($r$, left column) and the axial ($z$, right column) axes. The black solid, red dashed, and green dotted curves are calculated for $a_{bf}$ equals 0, 300$a_0$ and 600$a_0$, respectively.}
\label{fig:densities}
\end{figure*}

We start with the zero-temperature energy functional of the mixture, which is \cite{Imambekov2006bot}
\begin{equation}\label{eq:energy_functional}
\begin{split}
E=&\int d^3r\left[-\frac{\hbar^2}{2m_b}\psi^*\nabla^2\psi+V_b \psi^*\psi + \frac{1}{2}g_{bb}(\psi^{*}\psi)^2
\right.\\
&+\frac{1}{9}\frac{\hbar^2}{2m_f}(\nabla \sqrt{n_f})^2+V_f n_f+\frac{\hbar^2}{2m_f}\frac{3}{5}(6\pi^2)^{2/3}n_f^{5/3}\\
&\left. + g_{bf} n_b n_f \right].
\end{split}
\end{equation} 
Here $\psi$ is the order parameter of the BEC, $n_f$ is the number density of the fermions, $V_b$ and $V_f$ are the corresponding trapping potentials, $g_{bb}=4\pi\hbar^2a_{bb}/m_b$ and $g_{bf}=2\pi\hbar^2a_{bf}(m_b^{-1}+m_f^{-1})$ are the boson-boson and boson-fermion coupling constants. Considering the static case, we ignore the dynamic phase of $\psi$ and have $\psi=\psi^*=\sqrt{n_b}$, where $n_b$ is the BEC number density. Consequenctly we can replace $-\psi^*\nabla^2\psi$ in Eq.~\eqref{eq:energy_functional} with $(\nabla \sqrt{n_b})^2$. The term with $\nabla$ for bosons  arise from the kinetic energy and can be ignored in the Thomas-Fermi (TF) limit. The $\nabla$ term for fermions is the leading term from the density-gradient correction, which is much smaller than other terms under the relevant experimental conditions~\cite{Lous2018pti}.

In order to obtain the static solution, we minimize the energy functional with the steepest decent method (also known as the imaginary time evolution) \cite{Press2007book}. The evolution of the densities from step $j$ to $j+1$ follows
\begin{equation}\label{eq:E_functional_one_step}
\sqrt{n}_{j+1}=\sqrt{n}_{j}- \frac{\delta E}{\delta n}\sqrt{n}_{j}\Delta \tau,
\end{equation} where $\Delta\tau$ is the step size of evolution, and its value should be large to ensure fast converging while small enough to avoid numerical instability. We normalize atom numbers after each time step under the constraints
\begin{equation}
\begin{split}\label{eq:LagrangeEq_NConstraint}
N_b&=\int n_b d^3r,\\
N_f&=\int n_f d^3r.
\end{split}
\end{equation} 
The equilibrium solution of $n_b$ and $n_f$ is obtained when the algorithm converges. 


The zero-temperature density profiles at $a_{bf}=$0, 300, and 600$a_0$ are plotted in Fig.~\ref{fig:densities} for our typical experimental conditions \cite{Lous2018pti} with $N_b=1.5\times 10^4$ bosons and $N_f=1.5\times 10^5$ fermions. The radial center-of-mass (COM) trapping frequency is 171 Hz (300 Hz) for bosons (fermions), and the aspect ratio of the elongated trap is $7.55$. We observe that the fermions are depleted from the trap center by the increasing $a_{bf}$ to a moderate value slightly above 300$a_0$. With stronger repulsions, the BEC is further  squeezed to form a core that is surrounded by the fermions via a thin interface.

\subsection{Overlap function}
\begin{figure}
\centering
\includegraphics[scale=1]
{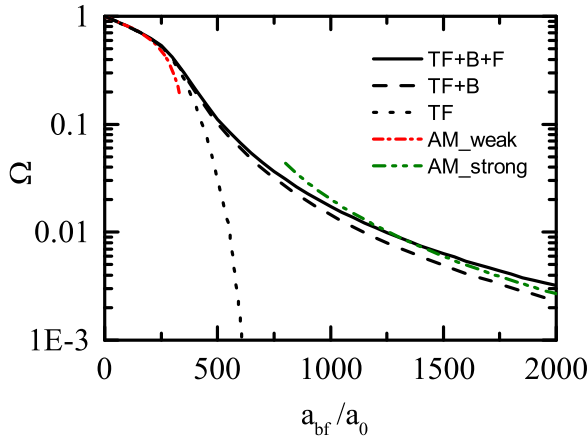} 
\caption{Comparison between the $\Omega$ from the numerical algorithms and the analytical models. The full numerical model produces the black solid curve (TF+B+T), applying TF approximation for both components in the numerical model gives the black dotted line (TF), and using the TF approximation only for the fermions leads to the black dashed curve (TF+B). The results from the analytical models for weak (AM\_weak, red dash-dot) and strong (AM\_strong, green dash-dot-dot) repulsive interactions are compared to the numerical results.}
\label{fig:overlap}
\end{figure}

We consider the normalized boson-boson-fermion (BBF) overlap at zero-temperature, because three-body recombination losses are dominated by BBF collisions and they can be used to probe the interface~\cite{Lous2018pti}. We ignore other three-body collisions involving two or more identical fermions, because they are suppressed at low temperatures by Pauli Blocking. Three-boson recombinations is also negligible in relevant experimental conditions, as the boson-boson scattering length is very small. The normalized BBF overlap is
\begin{equation}
\Omega=\frac{\int n_b^2n_fdV}{\int \tilde{n}_b^2\tilde{n}_fdV},
\end{equation} 
where $\tilde{n}$ denotes the density profile in the absence of Bose-Fermi interaction. The overlap function from the numerical results is plotted as the black solid line (TF+B+F) in Fig.~\ref{fig:overlap}. We observe a smooth decrease of $\Omega$ as $a_{bf}$ increases. 

Now we compare our full model with the Thomas-Fermi (TF) limit to study the influence from the kinetic energy of the BEC and the density gradient of the fermions. Within TF approximation, we remove $\nabla \sqrt{n_b}$ and $\nabla \sqrt{n_f}$ in Eq.~\eqref{eq:energy_functional} and obtain $\Omega$. The results are plotted as the dotted black line (TF) in Fig.~\ref{fig:overlap}. We observe that the overlap decreases drastically and vanishes near $a_{bf}=600a_0$, where phase separation takes place. 

We further test the importance of $\nabla \sqrt{n_f}$ by excluding it from the full model while keeping the BEC kinetic energy term. The results are shown as the dashed black curve (TF+B) in Fig.~\ref{fig:overlap}. We find the TF+B curve being very close to the full calculation, indicating the fact that the influence from the fermion density gradient is very weak. 

\subsection{Analytical results at small $a_{\rm bf}$}

We start from the mean-field TF density profiles in a harmonic trap as
\begin{align}
n_f&=n_{f0}\left(1-\frac{g_{bf} n_b}{\mu_{f0}} \right)^{3/2},\label{eq:nf_TF}\\
n_b&=n_{b0}\left(1-\frac{U_b+g_{bf} n_f}{ \mu_{b0}}\right),
\end{align} where $n_{f0}$ and $n_{b0}$ are the peak densities of the fermions and the bosons, $\mu_{f0}$ and $\mu_{b0}$ the global (position independent) chemical potentials, and $U_b$ the trapping potential of bosons. In Eq.~\eqref{eq:nf_TF}, we applied  the fermionic reservior approximation (FRA)~\cite{Lous2018pti} by ignoring any trapping potential $U_f$ for fermions and treat the fermions as a Fermi sea with a constant chemical potential. The FRA is valid as long as the BEC extents in a region much smaller than the fermion gas.  

We first consider the homogeneous case, where $U_b$ vanishes. In the weak interaction regime, we perform a Taylor expansion in $g_{bf}$ and find $\delta \mu_b/\delta n_b \approx g_{bb}+g_2$, where
\begin{equation}\label{eq:g2}
g_2=-\frac{3}{2}g_{bf}^2\frac{n_f}{\mu_f},
\end{equation}
and for the leading term we have $n_f\approx n_{f0}$ and $\mu_f\approx \mu_{f0}$. This correction $g_2$ to $g_{bb}$ indicates a fermion-mediated boson-boson interaction, which has been observed in Refs.~\cite{DeSalvo2019oof,Edri2020oos} and interpreted as the long-wavelength limit of the Ruderman-Kittel-Kasuya-Yosida (RKKY)~\cite{Ruderman1954iec} interaction. In our Eq.~\eqref{eq:g2} the scaling $g_2\propto g_{bf}^2$ means the mediated interaction is a second-order mean-field effect in the adiabatic limit. 

Now we consider trapped mixtures and the effects of a small $a_{bf}$ to the overlap factor $\Omega$. In this case, the fermions are slightly repelled from the trap center while the BEC is weakly compressed. Then the Taylor expansion of $\Omega$ on the small parameter $\eta=g_{bf}n_{b0}/\mu_{f0}$ leads to
\begin{equation}\label{eq:overlap_small_abf}
\Omega=1-\eta+\left(\frac{2}{11}+\frac{9d}{10}\right)\eta^2+\left(\frac{10}{429}-2d\right)\eta^3+O[\eta^4]
\end{equation}
where $d=\mu_{f0}n_{f0}/\mu_{b0} n_{b0}$ is on the order of 1 in our experimental conditions.
Because within the TF description the overlap function does not depend on the aspect ratio of the trap, we evaluate Eq.~\eqref{eq:overlap_small_abf} up to the third order and plot the result in Fig.~\ref{fig:overlap} as a red dash-dot curve. This analytical model agrees very well with the numerics in the weak interaction regime, and it only begins to deviate near the region of phase separation.
\subsection{The analytical model at large $a_{\rm bf}$}\label{sec:static_psl}
Let us discuss the opposite limit, where the $a_{bf}$ is very large and the fermions and bosons are separated. Obviously, the residual overlap in this regime is beyond the TF approximation, and we have to consider at least the kinetic energy of the BEC. We start from an infinite system without trapping potential, where we have only BEC (fermions) at $x\rightarrow\infty$ ($x\rightarrow-\infty$) and the components are separated by an interface parallel to the $y$-$z$ plane. 

By applying the TF approximation to fermions and the pressure balance at the interface (see App.~\ref{app:PSL_static}), we obtain the dimensionless differential equation for the order parameter $\psi =\sqrt{n_b}$ of the BEC as
\begin{equation}
\phi = -\frac{\partial^2 \phi}{\partial X^2}+\phi^3+\frac{5\eta}{4}\left(1-\phi^2 \eta\right)^{3/2}\phi,
\end{equation}
where $\psi$ is normalized by $\phi=\psi/\sqrt{\mu_{b0}/g_{bb}}$, and $x$ by the BEC healing length $\xi_{b0}=\hbar/\sqrt{2m_b\mu_{b0}}$ as $X=x/\xi_{b0}$.
The solution with boundary condition $\phi(+\infty)=1$ and $\phi(-\infty)=0$ is given by 
\begin{equation}
\frac{dX}{d\phi}=\sqrt{\frac{2}{(1-\phi^2)^2-(1-\phi^2 \eta)^{5/2}\Theta(1-\phi\sqrt{\eta})}},
\end{equation}
where $\Theta$ is the Heaviside step function.
The one-dimension BBF integral along the $x$-axis is then 
\begin{equation}
\begin{split}
I_{x}&= \frac{1}{2}\int_{-\infty}^{+\infty} n_b^2n_fdx\\
&=\frac{\sqrt{2}\xi_b }{2}n_{b0}^2 n_{f0}\eta^{-5/2}\\
&\times\int_0^1du \frac{ u^4 (1-u^2)^{3/2}}{\sqrt{(1-u^2/\eta)^2-(1-u^2)^{5/2}}}
\end{split} 
\end{equation}
where we used $u=\phi\sqrt{\eta}$ for simplification and the factor 1/2 results from the suppression of thermal bunching in a BEC involving two identical bosons (see also Eq.~\eqref{eq:omega_BFt}). We recognize that the overlap integral decreases proportional to $a_{bf}^{-5/2}$ at strong repulsive interactions. 

To calculate $\Omega$ for a trapped mixture in the phase separated regime, we consider that $\xi_b$ is much smaller than $R_b$ and $R_f$. Then the atom number conservation and the pressure balance at the interface fix the equilibrium condition of the system, e.g.\;$R_b$,  $R_f$ and the radial position of the interface $r=\zeta$ (see App.~\ref{app:PSL_static}). We evaluated $I_x$ at the interface and multiply it with the surface of the ellipsoidal interface, whose semi-axes are ($\zeta,\zeta,A \zeta$). We then normalize the outcome with the overlap integral at zero $a_{bf}$, which is obtained from the numerical model (TF+B) while an analytical approximation is also available \cite{Lous2018pti}. The resulted $\Omega$ is plotted in Fig.~\ref{fig:overlap} as the green dash-dot-dot curve, which is consistent with the numerical model. The small discrepancy between the numerical and  analytical results is probably caused by the surface tension of the interface \cite{Schaeybroeck2008ito} and the finite value of $\xi_b/R_b$ and $\xi_b/R_f$.

We also compare this result with a simple intuitive model, which assumes that the BEC is facing a hard wall at the interface and its $n_b(x)=n_{b0}\tanh^2({x}/{\sqrt{2}\xi_{b0}})$ is suppressed to zero within its healing length $\xi_b$. At the steep and rigid mean-field potential induced by the BEC, the Fermi sea is filled up to the Fermi energy. The corresponding $I_x$ agrees with the full calculation with a deviation less than 20\% when $\eta\gg 1$.

\subsection{The thermal faction of bosons}
\begin{figure*}
\centering
\includegraphics[scale=0.8]
{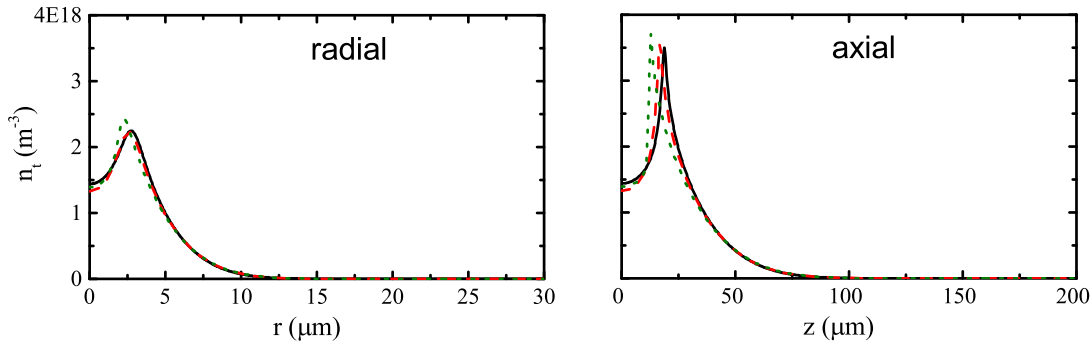} 
\caption{The density profiles of the thermal bosons ($n_t$) along the radial ($r$, left column) and the axial ($z$, right column) axes. The black solid, red dashed and green dotted curves are calculated for $a_{bf}$ equals 0$a_0$, 300$a_0$ and 600$a_0$, respectively. We have used the $n_b$ and $n_f$ in Fig.~\ref{fig:densities} to calculate the mean-field for thermal bosons.}
\label{fig:thermalBosonDensity}
\end{figure*}
In this model we assume $n_t$ to be influenced by $n_b$ and $n_f$, but not vice versa. This approximation is valid as long as we have $n_t\ll n_b$. We then calculate $n_t(r,z)$ with a Bose-Einstein distribution for thermal bosons, which is
\begin{equation}
n_t=\left(\frac{m_bk_B T}{2 \pi \hbar^2}\right)^{3/2}\text{Li}_{3/2}\left[e^{(\mu-U_b)/k_BT}\right],
\end{equation}
where the total effective potential $U_b$ for a thermal boson is $V_b+2 g_{bb}n_b+g_{bf}n_f$ and Li$_{3/2}$ is the polylogarithm function of order 3/2. The chemical potential $\mu_b$ of the BEC is obtained from the known $n_b$ and $n_f$ using $\mu_b \psi=(-\hbar^2\nabla^2/2m_b+g_{bb}n_b+g_{bf}n_f) \psi$ and $\psi=\sqrt{n_b}$. The temperature $T$ is fixed numerically to fulfil the atom number constraint $N_t=\int n_t dV$, and we finally get the density distribution $n_t$, which is shown in Fig.~\ref{fig:thermalBosonDensity}. Here the total boson number is $3\times10^4$ and the BEC fraction is 50\%. We notice that the thermal bosons form a shell-like structure at the edge of the BEC, because both the BEC and the fermions are repulsive to thermal bosons.

Although the density $n_t$ of the thermal component of the boson gas is typically almost two order of magnitude lower than the BEC density $n_b$, the significant thermal fraction (up to $\sim50\%$) of the bosons in typical experimental conditions leads to a correction to the total overlap between bosons and fermions. The generalized overlap $\Omega$ including the thermal bosons is defined as \cite{Lous2018pti}
\begin{equation}\label{eq:omega_BFt}
\Omega=\frac{\int (\frac{1}{2}n_b^2n_f+n_b n_t n_f+n_t^2n_f)dV}{\int (\frac{1}{2}\tilde{n}_b^2\tilde{n}_f+\tilde{n}_b\tilde{n}_t\tilde{n}_f+\tilde{n}_t^2\tilde{n}_f)dV},
\end{equation} 
where $n_t$ and $\tilde{n}_t$ are the thermal boson density with and without the Bose-Fermi interaction.

\begin{figure}[t]
\centering
\includegraphics[scale=1]
{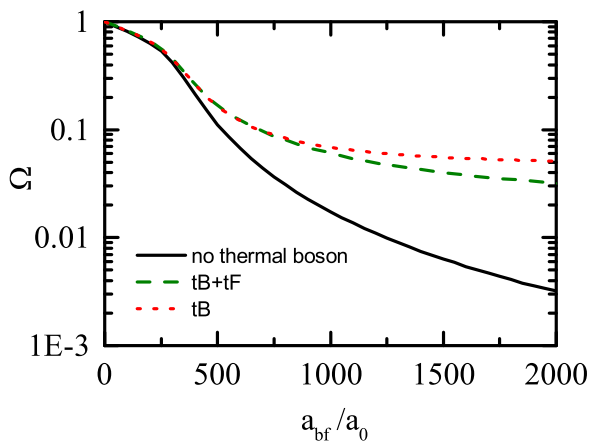} 
\caption{Comparison between the overlap factor $\Omega$ from the numerical model with and without including thermal bosons. The black solid line is identical to the one in Fig.~\ref{fig:overlap}, which ignores the thermal bosons. The green dashed and red dotted curves are results including thermal bosons, which are interacting either with both the BEC and the fermions (tB+tF) or with only the BEC (tB).}
\label{fig:overlap2}
\end{figure}

We calculate $\Omega$ from $n_b$, $n_f$ and $n_t$, and plot the results as the green dashed curve in Fig.~\ref{fig:overlap2}. The black solid curve in Fig.~\ref{fig:overlap2} excludes thermal bosons and is identical to the one in Fig.~\ref{fig:overlap}. If we ignore the interaction between the thermal bosons and the fermions, $\Omega$ become the red dotted curve in Fig.~\ref{fig:overlap2}. We find that the residual of $\Omega$ in the strong interaction regime has a value of a few percent, and it slowly decreases when the repulsion between the thermal bosons and the fermions is considered. 

\subsection{Finite-temperature effects on fermions}
At a finite temperature the BEC is only weakly influenced by the thermal bosons, therefore we can analyze the degenerate and non-degenerate parts separately. But we cannot define a thermal or degenerate part of the fermionic gas since all the single particle orbitals are correlated. Consequently we estimate the finite temperature effects of fermions in a perturbative way, i.e.\;calculating the finite-$T$ fermion density $n_{f}$ from the $n_b$ known at zero-$T$ with a Fermi-Dirac distribution. We will discuss only the temperature effects on the overlap integral, i.e.\;the numerator in Eq.~\eqref{eq:omega_BFt}, because the denominator is fixed in our  definition of $\Omega_{\rm eff}$.

With a typical temperature of $T/T_f\approx0.12$, $n_{f}$ is about $5\%$ lower than the zero-$T$ solution at the trap center and lightly spreads out at the edge, e.g.~$r=R_f$ in the radial direction. Since our experiments have $R_f/R_b\approx8$, we ignore the inhomogeneity of $n_f$ in the BEC region and expect about $5\%$ down shift of the overlap integral.
On the other hand, the thermal boson cloud extends much wider than the BEC and is much less sensitive to the finite-$T$ correction of $n_f$. 

Another effect, which also exists at zero temperature, is the unitary limit of the cross section of the Bose-Fermi scattering. The fermions have a kinetic energy on the order of the Fermi energy $E_F$, which is much larger than the kinetic energy of the bosons and leads to a reduction of the Bose-Fermi cross section. In the relevant experimental conditions we have $1/k_F\approx 4500a_0$, where $E_F=\hbar^2k_F^2/2m_f$. Since the mean kinetic energy of fermions at the trap center is $3E_F/5$, we estimate that the cross section scales proportional to $a^2/(1+0.6\times a^2k_F^2)$. Therefore the reduction of the Bose-Fermi cross section is negligible in typical experimental conditions, where $a_{bf}$ is always below about $2000a_0$.

\section{radial breathing mode}\label{sec:dynamic}
In our recent experiment on an elongated Bose-Fermi mixture~\cite{Huang2019bmo}, we have observed 
a significant frequency shift of the radial breathing mode (RBM) of the BEC when phase separation takes place. 
This motivates us to theoretically investigate the collective mode of a BEC immersed in a large Fermi sea at various values of $a_{bf}$. To model the dynamics of the BEC, we use the Gross-Pitaevskii equation (GPE) and include a mean-field potential $g_{bf} n_f$ formed by the fermions. For the dynamics of fermions, we utilize different models. The first model we introduce here is the adiabatic Fermi sea (AFS) model, which assumes that the fermions adapt adiabatically to the perturbations of the BEC. This model can be solved by either performing time evolutions (TE) or extracting eigenvalues (EG) from linearized equations. Then in the second model, we use the collisionless Boltzmann-Vlasov equation (BVE) to describe the fermions. The BVE is solved numerically with the test-particle method (TPM), while an analytical result is also achieved at the phase separation limit (PSL). We obtain the RBM frequency $\omega$ from different models and compare the results.

In our following calculations we approximate our elongated mixture to an axially invariant system with cylindrical symmetry. The radial plane of the model corresponds to the radial plane of the  mixture at the trap center. The static density profiles of the mixture in the radial plane are obtained from the numerical model described in Sect. \ref{sec:static_numeric}. The temperature of the system is assumed to be zero in our oscillation models. 

\begin{figure}[h]
\centering
\includegraphics[width=\columnwidth]
{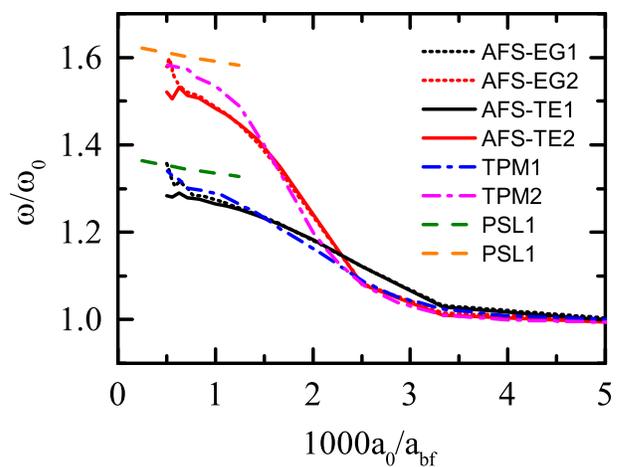} 
\caption{The radial breathing mode frequency shift $\omega/\omega_0$ calculated from different models and parameters. Two sets of atom numbers are used here: Set 1 with $N_b=1.6\times10^4$ and $N_f=1.03\times10^5$ and Set 2 with  $N_b=8\times10^3$ and $N_f=1.68\times10^5$. The black (red) solid curve is given by the AFS approximation by the direct time evolution (AFS TE) method for Set 1 (Set 2). The black (red) dotted curves also includes the AFS approximation but is obtained by finding the eigen values of the linearised equations (AFS EG) for Set 1 (Set 2). The blue and pink dash-dot curves are the results of the GPE-BVE mode solved by the TPM for two sets of parameters. The outcome of the PSL model at large $a_{bf}$ values are shown as the green and yellow dashed curves for the two sets.}\label{fig:freq_shift}
\end{figure}

\subsection{Adiabatic Fermi sea approximation}\label{sec:AFSA}
If the density of fermions is high enough to support a Fermi velocity $v_F=\sqrt{2E_F/m_f}$ much larger than the sound velocity $v_s=\sqrt{g_{bb}n_b/m_b}$ of the BEC, the fermions  follow adiabatically the fluctuations of the BEC density within a mean-field picture \cite{Yip2001cmi}. This adiabatic Fermi sea (AFS) approximation is valid in relevant experimental conditions as long as the fermions are not depleted from the BEC region. 

We formally write down the time-evolution equation of the BEC dressed by the adiabatic Fermi sea as 
\begin{equation}\label{eq:time_evolve}
i \hbar \frac{\partial \psi}{\partial t}=-\left[\frac{\hbar^2}{2m_b}\nabla^2+V_b +g_{bb}|\psi|^2+g_{bf}n_f\right] \psi.
\end{equation}
The time-dependent fermion density $n_f=C_f(\mu_{f0}-g_{bf}|\psi|^2)^{3/2}\Theta(\mu_{f0}-g_{bf}|\psi|^2)$, where $\mu_{f0}$ is the global Fermi energy and $C_f=(2m_f/\hbar^2)^{3/2}/6\pi^2$, is calculated with the TF approximation and FRA of the fermions. 

In order to numerically solve Eq.~\eqref{eq:time_evolve}, we approximate our cigar-shaped cloud with a cylindrical system, whose radial plane represents the radial plane of the mixture at the trap center. We describe $\psi$ in the radial plane by setting up a one-dimensional complex-valued grid for $\psi$ along the radial direction.

Our first way to solve Eq.~\eqref{eq:time_evolve} is numerically calculating the time evolution of $\psi$. Inspired by our experiment~\cite{Huang2019bmo}, where we excited the RBM by switching $a_{bf}$ between a small and a large value, we perform a similar process in our simulations for a given $a_{bf}$. We take the static solution $\psi_0$ at $a_{bf}-100a_0$ as the initial value of $\psi$. Then we switch to $a_{bf}$ and let $\psi$ evolve in time according to Eq.~\eqref{eq:time_evolve}, while the TF density $n_f$ at each time-step is calculated from $|\psi|^2$ at that moment. We record the averaged BEC width $\left\langle r\right\rangle=\int dr  r W(r)/\int dr W(r)$, where the weight function is $W(r)=2\pi r n(r)$, as a function of time and fit it to a cosine function to extract the oscillation frequency $\omega$. 

The calculated dependence of $\omega$ on the interaction strength is plotted as solid curves in Fig.~\ref{fig:freq_shift} and marked as AFS-TE for the AFS approximation and the time-evolution method. The black solid curve corresponds to an experimental setting with a boson number of $N_b=1.6\times 10^4$ and a fermion number of $N_f=1.03\times 10^5$ (named Set 1) while the red solid curve uses $N_b=8\times 10^3$ and $N_f=1.68\times 10^5$ (Set 2). These parameters correspond to our experimental conditions~\cite{Huang2019bmo}. We use $x=1000a_0/a_{bf}$ for the horizontal axis in the plot, and the y-axis is normalized to the RBM frequency $\omega_0$ at $a_{bf}=0$. The AFS-TE model shows that the RBM frequency $\omega$ barely changes at small scattering lengths. Then it starts to increase rapidly near $x=3$, where the fermions are depleted from the trap center. Finally $\omega$ tends to reach a maximum frequency shift near around $x=1$. The numerics begin to fail when the mixture is deeply in the phase separation regime and the interface depth is comparable to the grid step-size.

\begin{figure}[t]
\centering
\includegraphics[width=\columnwidth]
{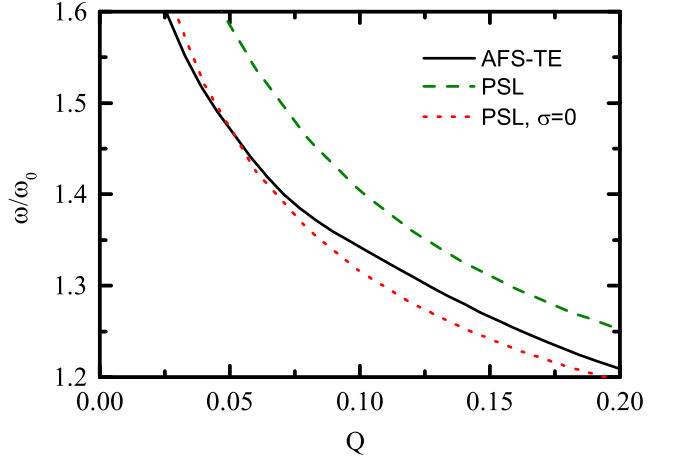} 
\caption{The RBM frequency $\omega/\omega_0$ at the PSL as a function of the boson number fraction $Q=N_b/(N_b+N_f)$. The black solid curve is calculated from the AFS-TE model. The PSL model with and without ($\sigma=0$) the surface tension effects give the green dashed and red dotted curves, respectively. We use a total atom number of $1.5\times 10^4$ and $a_{bf}=1300a_0$.}\label{fig:freq_q}
\end{figure}

We notice that the maximum shift of $\omega$ is higher when $N_f/N_b$ is larger. So we define the boson number fraction $Q=N_b/(N_b+N_f)$, fix the total atom number $N_b+N_f=1.5\times 10^5$, and calculate the RBM frequency at $a_{bf}=1300a_0$ for various $Q$ values. The results are presented as the black curves in Fig.~\ref{fig:freq_q} for a range of $Q$ that is reachable in the experiment. We find that $\omega/\omega_0$ increases faster when $Q$ is close to 0. 

As an alternative numerical method, we linearize and solve Eq.~\eqref{eq:time_evolve} for small perturbation $\delta\psi$ and $\delta n_f=-(3/2)g_{bf} C_f^{2/3} n_f^{1/3} \delta (\psi^* \psi)$.
We take the ansatz $\psi=(\psi_0+ue^{-i\omega t}+v^*e^{i\omega t})e^{-i\mu_b t/\hbar}$ near the equilibrium solution $\psi_0=\psi_0^*=\sqrt{n_{b}}$ 
and arrive at the linearized effective GPE as
\begin{equation}\label{eq:GPE_linear}
\begin{split}
\hbar \omega u=&\left[H_{b}+(2g_{bb}+g_2)n_b\right]u+(g_{bb}+g_2)n_b v\\
-\hbar \omega v=&\left[H_{b}+(2g_{bb}+g_2)n_b\right]v+(g_{bb}+g_2)n_b u\\
\end{split}
\end{equation}
where $H_{b}=-(\hbar^2/2m_b)\nabla^2+V_b+g_{bf}n_f-\mu_b$ accounts for the kinetic energy, the boson trapping potential, the mean-field potential induced by a static Fermi, and the global chemical potential $\mu_b$ of BEC obtained from the static solution. The boson-boson interaction that is mediated by the Fermi sea is given in Eq.~\eqref{eq:g2}.

In order to numerically solve Eq.~\eqref{eq:GPE_linear} for a cylindrical system, we discretize $u(r)$ and $v(r)$ along the radial direction.
With the $n_b$ and $n_f$ obtained in Sec.~\ref{sec:static_numeric}, we obtain the eigenfrequency $\omega$ by diagonalizing Eq.~\eqref{eq:GPE_linear} in its matrix form with a regularized boundary condition. The calculated $\omega/\omega_0$ values of the lowest RBM are shown in Fig.~\ref{fig:freq_shift} as the black and red dotted curves (AFS-EG1 and AFS-EG2) for the two sets of atom numbers. Although the diagonalization method requires $n_f>0$ and becomes no longer fully valid when $a_{bf}$ is so large that fermions are depleted from the trap center, we find its results agree very well with the outcomes of the direct time evolution before numerical instabilities take over in the deeply phase-separated regime. 

Finally, we discuss our Eq.~\eqref{eq:GPE_linear} in comparison with other experimental~\cite{DeSalvo2019oof} and theoretical works (e.g.~\cite{Tsurumi2000dom,Santamore2008fmi,De2014fml}) on weakly interacting Bose-Fermi mixtures. In the earlier works, a simple replacement of the boson coupling constant $g_{bb}\to g_{bb}+g_2$ transforms the dynamics from a pure BEC to a weakly interacting Bose-Fermi mixture. In our model, such a transformation is not explicit in Eq.~\eqref{eq:GPE_linear} because of the terms proportional to $2g_{bb}+g_2$. Our results recover $g_{bb}\to g_{bb}+g_2$ only if we take the limit of $g_{bf} n_b\ll \mu_f$, where $g_{bf}\delta n_f\approx g_2n_b$ is valid.
Moreover, we find a small $g_{bf}$ also induces a correction (buoyancy-like effect) $-\frac{3n_f V_f}{2\mu_f V_b}g_{bf}$ to the boson potential $V_b$ when a weak fermion potential $V_f\ll \mu_f$ is considered. In general, we expect our results to be valid also beyond the weak-interaction regime.


\begin{figure*}[th]
\centering
\includegraphics[width=2\columnwidth]
{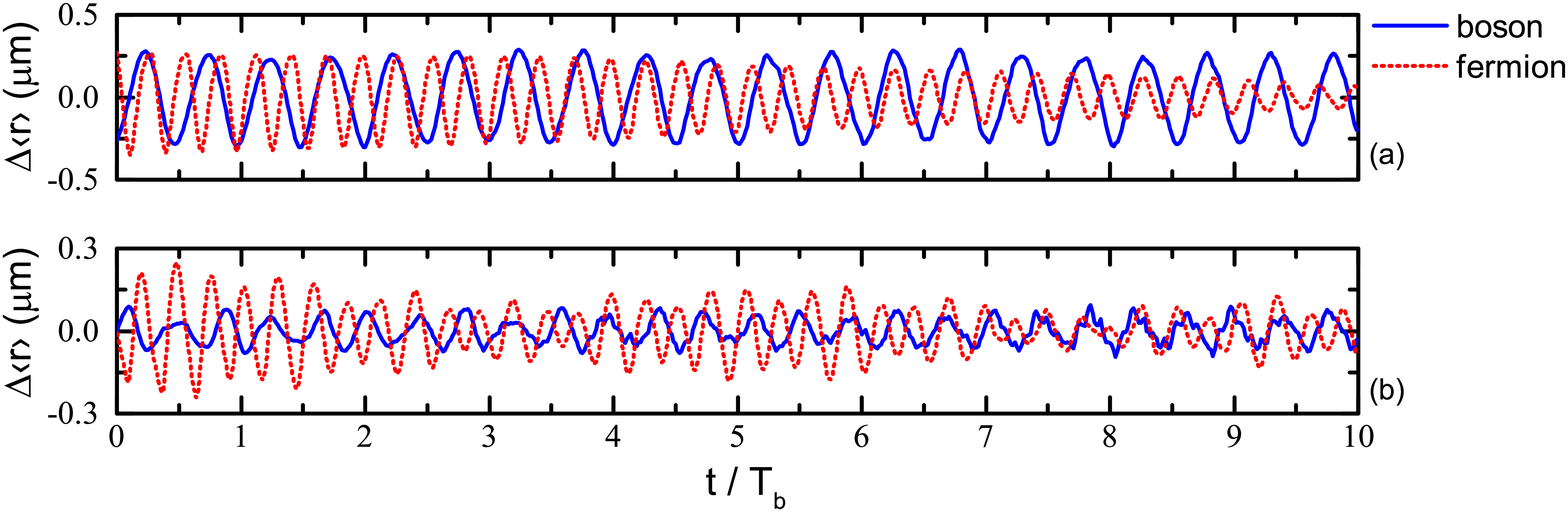} 
\caption{Simulated width evolution of the bosonic (blue solid) and the fermionic (red dotted) cloud, where the GPE-BVE model is used. The time $t$ is normalized to the single boson COM period $T_b$ in the trap.  The $y$-axis shows deviation of the density-averaged width $\langle r \rangle$ of the cloud from its mean value. The oscillation is excited with 3 consecutive quenches of the Bose-Fermi interaction (scheme Exct1). The panel (a) presents the oscillation at $a_{bf}=100a_0$, and panel (b) corresponds to $a_{bf}=1000a_0$.}\label{fig:bve_osci_3kick}
\end{figure*}
\subsection{Kinetic equation simulation for fermions}

To describe the dynamics of the degenerate single component Fermi gas by first principles, we utilize the Boltzmann-Vlasov equation (BVE) with vanishing collisions, i.e.
\begin{equation}\label{eq:BVE}
\partial_t f+\frac{1}{m_f}\vec{p}\cdot\nabla_r f- \frac{1}{m_f}\vec{F}_f\cdot \nabla_v f=0,
\end{equation}
where $f(\vec{p},\vec{r})$ is the fermion distribution function in the phase space, $m_f$ the mass of a single fermion. The force $\vec{F}_f$ on the fermions is given by the trapping potential $V_f(r)=m_f \omega_f^2r^2/2$ and the repulsion $g_{bf}n_b$ from the BEC. 

Equation~\eqref{eq:BVE} can be solved numerically with the quasi-particle method \cite{Maruyama2005moa,Brack1997sp}, which uses a cloud of $\tilde{N}$ classical pseudo particles (test particles) with mass $m_f$ to sample the phase space density of $N$ real fermions. The kinetic equation is then simulated with the Newtonian equations of the test particles in the external potential, and the real fermion density is calculated from the test particle density with a scaling factor $N/\tilde{N}$.

In our calculations for a cylindrical system, we implement a one-dimensional spatial grid for $\psi$ and take a value of $\tilde{N}$ so that the shot noise of $n_f$ at the grid points is smaller than the thermal statistic noise at $T/T_F=0.1$. 
In order to excite the collective mode, we start with the static densities at $a_{bf}=0$ and apply three consecutive quenches of the scattering length $a_{bf}$ between 0 and $700a_0$, which closely imitates the relevant experimental sequence~\cite{Huang2019bmo}. 
After the excitation stage, the time evolution of the effective width $\langle r\rangle$ for the BEC and the fermions are recorded to extra the frequency. 

\begin{figure}[h]
\centering
\includegraphics[width=\columnwidth]
{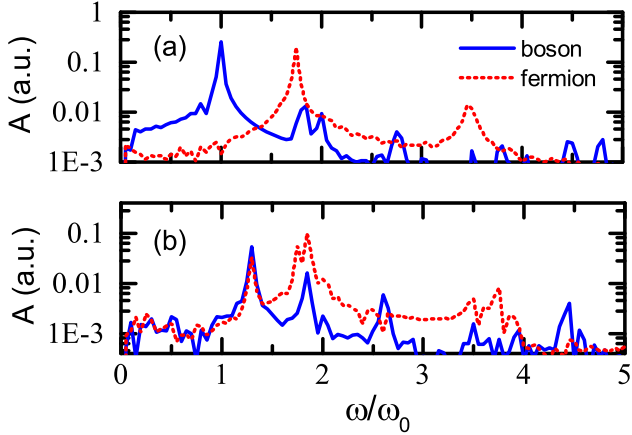} 
\caption{The FFT spectrum amplitude $A$ of the oscillations shown in Fig.~\ref{fig:bve_osci_3kick}. The radial breathing mode frequency $\omega$ is normalized to its reference value $\omega_0$ at $a_{bf}=0a_0$. The panel (a) and (b) shows the results from $a_{bf}=100a_0$ and $1000a_0$ respectively.}\label{fig:bve_fft}
\end{figure}

The typical time evolution of the $\langle r\rangle$ of the BEC (the blue solid curve) and the fermions (the red dotted curve) are plotted in Fig.~\ref{fig:bve_osci_3kick}. Our calculation uses atom numbers from Set 1 (see Sec.~\ref{sec:AFSA}). And $a_{bf}=100a_0$ and $a_{bf}=1000a_0$ applies to panel (a) and (b) in Fig.~\ref{fig:bve_osci_3kick}, respectively. We recognize that the oscillation is not necessarily a simple sinusoidal wave, and it contains  possibly multiple frequency components. So we apply the fast-Fourier-transformation (FFT) to the data and obtain the power spectra, which are shown in Fig.~\ref{fig:bve_fft}. 

We find that two frequency components  are important in the spectra, one is close to or slightly above $\omega/\omega_0=1$ and another is close to $\omega/\omega_0=1.8$. Keep in mind that $\omega_{f}/\omega_b=1.75$ and the BEC extents much narrower than the fermionic gas, we recognize the higher frequency component as the single fermion mode, which is only slightly changed by the small BEC. On the other hand, the main peak in the  BEC spectrum indicates that the BEC RBM is coupled to the fermions and forms a lower branch of the oscillating spectrum.


To be consistent with former analysis on the frequency within this work, we fit the oscillation of width with a cosine function and extract the frequency $\omega$. The results are plotted in Fig.~\ref{fig:freq_shift}. The blue (pink) dash-dot line shows the results for Set 1 (Set 2). The results from the TPM are consistent with the AFS model. Furthermore, we also vary the initial excitations, e.g.\;the amplitude of the oscillation, and extract $\omega$ in different time sections of the evolution. And we observe only minor changes (less than about 5$\%$) of the value of $\omega$.

\subsection{Analytical model of oscillation at the phase separation limit (PSL)}
At the phase separation limit (PSL), both the static and dynamic properties of the Bose-Fermi mixture can be studied analytically \cite{Schaeybroeck2008ito,Schaeybroeck2009tps}. Following the procedure introduced in Ref.~\cite{Schaeybroeck2009tps} for a spherical case, we investigate the RBM of the BEC in an infinitely-long cylindrical mixture, which is more relevant to experiments. Starting from the TF approximation for both components, we assume that the two components in the mixture is connected by an infinitely thin interface layer, which has a surface tension effect with coefficient $\sigma$ \cite{Schaeybroeck2008ito} (see also App. \ref{app:surface_tension}). We first obtain separately the formal solution of the BEC density and the fermion phase space distribution function $f(r,p)$, then solve the problem of the mixture by matching the boundary conditions, i.e.~flux and pressure, at the interface. The equilibrium solution of the densities in the mixture has been used in Sec.~\ref{sec:static_psl} and discussed in details in App.~\ref{app:PSL_static}. We will now find the oscillation frequency $\omega$ of the RBM in the mixture.

The collective modes of a trapped BEC in the TF limit are well understood~\cite{Pitaevskii2016book}. The radial mode of a BEC with zero angular momentum in a cylindrical system corresponds to a density fluctuation of 
\begin{equation}\label{eq:2F1_cyl}
\begin{split}
&\delta n_b(r)\propto\\
&F\left(\frac{1+\sqrt{1+2\frac{\omega^2}{\omega_b^2}}}{2},\frac{1-\sqrt{1+2\frac{\omega^2}{\omega_b^2}}}{2},1,\frac{r^2}{R_b^2}\right),
\end{split}
\end{equation}
where $F$ is the hypergeometric function ${_2F_1}$, $\omega$ the frequency of the collective mode, $\omega_b$ the COM trapping frequency of bosons, and $R_b$ the Thomas-Fermi radius of the BEC.

At the interface $r=\zeta$ of the mixture there is no exchange of components. This means that the velocity of the BEC is equal to the speed $\dot{\zeta}$ of the moving interface. Consequently we have 
\begin{equation}\label{eq:BC_BEC}
{\delta\zeta}=\frac{g_{bb}}{\omega^2m_b }\partial_r\delta n_b,
\end{equation} where we denote $\partial_r n_b=\partial n_b/\partial r$ for simplicity.

In the phase separation limit, the kinetic equation for fermions reduces to  
\begin{equation}\label{eq:BVE_PS}
\partial_t f+\vec{v}\cdot\nabla_r f-\omega_f^2 \vec{r}\cdot \nabla_v f=0.
\end{equation}
We then apply the ansatz describing the deformation of the Fermi surface as $f=f_0+\delta(|\vec{p}|-p_f)u(r,\alpha,\beta)e^{-i\omega t}$, where $\alpha=\cos\phi$ with $\phi$ the between the momentum $\vec{p}$ and the position $\vec{r}$ in the radial plane, $\beta=\cos\theta$ with $\theta$ the angle between $\vec{p}$ and the longitudinal $z$-axis, and $p_f(r)=\sqrt{2 m_f\mu_f(r)}$ is the local Fermi momentum. Then we get the linearised kinetic equation
\begin{equation}
\begin{split}
-i\omega u +\omega_f\rho\alpha\sqrt{1-\beta^2}\partial_r u 
+\omega_f(1-\alpha^2)g(r,\beta)\partial_\alpha u
\\+\omega_f\alpha\beta\sqrt{1-\beta^2}\frac{r}{\rho}\partial_{\beta}u=0,
\end{split}
\end{equation} where $\rho=(R_f^2-r^2)^{1/2}$ with $R_f$ the Thomas-Fermi radius of fermions and $g(r,\beta)=(\rho\sqrt{1-\beta^2})/r-r/(\rho\sqrt{1-\beta^2})$.
The corresponding solution has the form 
\begin{equation}\label{eq:BVE_cyl_u_form}
u(r,\chi)=\mathcal{F}[r^2(R_f^2-r^2)(1-\alpha^2)(1-\beta^2)]e^{-i\omega \tau/2},
\end{equation}
where $\mathcal{F}[x]$ is an arbitrary function of $x$ and 
\begin{equation}\label{eq:tau_cyl}
\tau(\zeta,\alpha,\beta)=\frac{\psi_0-\arctan[2\alpha/g(\zeta,\beta)]}{\omega_f},
\end{equation}
with $\psi_0=\pi\Theta[g(r,\beta)]$. As $\tau$ characterizes the phase of the oscillating system, we recognize it as the time for a single fermion with Fermi velocity $v_F$ to depart and then return to the interface. Equation~\eqref{eq:tau_cyl} is valid when $\alpha\in[0,1]$ and beyond that we have $\tau(-\alpha,\beta,r)=-\tau(\alpha,\beta,r)$. It is also obvious that $\tau(\alpha,-\beta,r)=\tau(\alpha,\beta,r)$.

Now we consider the non-penetration boundary condition for fermions at the interface. It requires that 
\begin{equation}
\left[u(\zeta,\alpha,\beta)-u(\zeta,-\alpha,\beta)\right]e^{-i\omega t}=2 m_f \alpha\sqrt{1-\beta^2} \dot{\zeta},
\end{equation} where $\dot{\zeta}=\partial_t \zeta=-i\omega\delta\zeta e^{-i\omega t}$ is the velocity of the phase boundary. Taking this into Eq.~\eqref{eq:BVE_cyl_u_form} and \eqref{eq:tau_cyl}, the fermion perturbation is solved to be
\begin{equation}\label{eq:u_cyl}
u(\zeta,\alpha,\beta)=2m_f\alpha\sqrt{1-\beta^2}(-i\omega\delta\zeta)(1-e^{i\omega\tau})^{-1}.
\end{equation}
We may formally expand $(1-e^{i\omega\tau})^{-1}=\sum_{n=0}^\infty e^{in\omega\tau}$ and recognize that this solution $u$ is constituted of a series of elementary excitations happened at earlier times.

The condition of pressure equilibrium  at the interface is $P_b-P_f=\sigma/\zeta$, where the boson pressure is $P_b=g_{bb}n_b^2/2$ and the pressure $P_f$ of the collisionless fermions at the interface is given by the corresponding momentum flux $\Pi(r)=(1/m_f h^3)\int d^3\vec{p} \alpha^2 (1-\beta^2) p^2 f(\vec{p},r)$ in the radial direction \cite{Schaeybroeck2009tps}. Together with the boson boundary condition Eq.~\eqref{eq:BC_BEC} we obtain the equation for the oscillation frequency $\omega$ as 
\begin{equation}\label{eq:BC_PSL_cyl}
\frac{\partial_r F}{F}=\frac{\omega^2 m_b\left(n_b-\frac{1}{\zeta}\frac{\partial \sigma}{\partial\mu_b}\right)}{-\frac{\sigma}{\zeta^2}+\frac{1}{\zeta}\frac{\partial \sigma}{\partial\zeta}+\frac{p_f^4 C_{\Pi}}{(2\pi\hbar)^3}-\partial_r(P_b-P_f)},
\end{equation}
where $C_{\Pi}=8\omega\int_0^{\pi/2} d\phi \int_{0}^{1} d\beta (1-\beta^2)^{3/2}\cos^3\phi
\cot\left({\omega\tau}/{2}\right)$ and all values are calculated at the interface~(see App.~\ref{app:osci_dp} for details).

Taking typical parameters from the  experiments~\cite{Huang2019bmo}, we obtain  $\omega$ and show the results as the green and orange dashed curves in Fig.~\ref{fig:freq_shift} for the two sets of atom numbers. We find the results from the PSL model is consistent with other models. The PSL model predicts a slow increase of $\omega$ at the PSL,  and gives slightly higher $\omega$ values than other models. The $Q$ dependence of the RBM frequency shift is presented in Fig.~\ref{fig:freq_q} as the green dashed curve. In order to check the influence from the surface tension effects, we also calculate $\omega$ with $\sigma=0$ and plot the outcomes as the red dash-dot curve in Fig.~\ref{fig:freq_q}. The PSL model without surface tension leads to a slightly lower frequency shift.

\section{Conclusions}\label{sec:conclusions}
We thoroughly investigated the static density profiles and the 
radial breathing mode in an elongated degenerate Bose-Fermi mixture. The experimentally relevant conditions of~\cite{Lous2018pti,Huang2019bmo}, under which the size of the BEC is much smaller than the Fermi sea, have been considered. We first presented in details our mean-field numerical model, which explains the smoothing of phase separation with the kinetic energy of the BEC~\cite{Huang2019bmo}. Then we obtained analytical forms of the density profiles in the limit of weak and strong repulsion. In particular, the latter shows intuitively the structure of the interface: The BEC at the interface behaves like being blocked by a wall potential, and the fermions penetrates into the mean-field potential of the BEC edge with a depth determined by the chemical potential of the Fermi sea. 

We presented three zero-temperature models for the RBM of the BEC in cylindrical mixtures, of which the results are consistent. We first describe the adiabatic Fermi sea model, which explained the significant shift of the RBM frequency observed across the phase-separation of the mixture~\cite{Huang2019bmo}. Within this model, we find the fermion-mediated interaction between bosons to be an adiabatic second-order mean-field effect, which is valid also beyond the weak-interaction regime under relevant experimental conditions. 
For very large $a_{bf}$ values, another full phase-separation model considering single-fermion trajectories gives results similar to those from the AFS model. Finally, we perform test-particle simulations for the RBM, and the outcomes are almost identical to that of the AFS model for a large range of repulsion strength. 

The remaining discrepancy between the observed RBM frequencies~\cite{Huang2019bmo} and the results from our zero-$T$ models, especially in the regime of full phase-separation, stimulates future experiments in more deeply cooled samples with further reduced imperfections, e.g.~anharmonicity of the trap. Comprehensive theories  including finite-$T$ effects~\cite{Liu2003fte,Grochowski2020bmo}, quantum fluctuations, and possible hydrodynamic properties at the interface are encouraged to study relevant experimental situations. 

\begin{acknowledgments}
We thank B. Van Schaeybroeck, A. Lazarides and T. Maruyama for intensive discussions and providing details, especially numerical results, of their earlier works. We acknowledge and thanks the valuable discussions with M. Baranov, D. Yang, R. Bijnen, S. Watabe about the theoretical models. We acknowledge and thank the comments on the manuscript from our experimental group R. Grimm, J. Walraven, I. Fritsche, R. Lous, C. Baroni, T. W. Grogan and E. Kirilov. This work was supported by the Austrian Science Fund FWF within the Spezialforschungsbereich FoQuS (F4004-N23)
and partially the project P32153-N36.

\end{acknowledgments}

\appendix

\section{Surface tension at the interface}\label{app:surface_tension}
In the phase separation limit (PSL), the surface tension coefficient of a flat boson-fermion interface is~\cite{Schaeybroeck2008ito,Schaeybroeck2009tps}
\begin{equation}\label{eq:sigma}
\sigma(\kappa)=\frac{\hbar\mu_b^{3/2}}{g_{bb}\sqrt{m_b}}G(\kappa),
\end{equation}
where $\mu_b$ is the chemical potential of BEC, $G(\kappa)=\int_0^1 dx \sqrt{(1-x^2)^2-(1-x^2/\kappa^2)^{5/2}\Theta(\kappa-x)}$, and $\kappa=\sqrt{g_{bb} \mu_f/g_{bf}\mu_b}$.

Considering a mixture in a harmonic trap with boson COM frequency $\omega_b$ we have
\begin{align}
\frac{\partial\sigma}{\partial \mu_b}&= \frac{\sigma}{\mu_b}\left(\frac{3}{2}-\frac{\kappa G'(\kappa)}{2}\right)=\frac{\sigma}{\mu_b}C_\sigma\\
\frac{\partial\sigma}{\partial r}&= \frac{r \sigma}{r^2-R_b^2}\left(\frac{3}{2}-\frac{\kappa G'(\kappa)}{2}\right)=-\frac{m_b\omega_b^2 }{2 \mu_b}r \sigma C_\sigma,
\end{align}
where $G'$ is the derivative of the numerical function $G$, and  $C_\sigma={3}/{2}-{\kappa G'(\kappa)/}{2}$. Here we used the fermionic reservoir approximation, which ignores the dependence of $\mu_f$ on $r$.

\section{Static density profiles at the PSL} \label{app:PSL_static}

Within the TF approximation, the density profiles of a phase separated Bose-Fermi mixture can be obtained analytically \cite{Schaeybroeck2008ito,Schaeybroeck2009tps}.
The conservation of atom numbers and the pressure balance at the interface require that
\begin{align}
\frac{N_b}{A}&=\frac{R_b^5(m_b \omega_b /\hbar)^2}{2a_{bb}}\left(\frac{\zeta^3}{3R_b^3}-\frac{\zeta^5}{5R_b^5}\right),\label{eq:Nb}\\
\frac{N_f}{A}&=\frac{R_f^6( m_f \omega_f /\hbar)^3}{72\pi}\left(\frac{3\pi}{2}-K(\zeta/R_f)\right),\label{eq:Nf}
\end{align}
where $A$ is the aspect ratio of the trap, $N_b$ and $N_f$ the boson and fermion number, and $K(x)=x(1-x^2)^{1/2}(14x^2-3-8x^4)+3\arcsin(x)$. The interface position $\zeta$, TF radii $R_b$ and $R_f$ of the BEC and the fermions are all given in the radial plane.

The pressure balance at the interface is 
\begin{equation}
P_b-P_f=c_\sigma\sigma/\zeta,\label{eq:pressure}
\end{equation}
where the pressure of the BEC is $P_b=g_{bb}n_b^2/2$ with $n_b=m_b\omega_b^2(R_b^2-\zeta^2)/2$, and the pressure of fermions is $P_f=(2/5)C_f\mu_f^{5/2}$ with $\mu_f=m_b\omega_b^2(R_f^2-\zeta^2)/2$ and $C_f=(2m_f/\hbar^2)^{3/2}/6\pi^2$. The coefficient $c_\sigma$ is 2 for a spherical system and 1 for a cylindrical system. Since $c_\sigma$ is not a constant on the interface of an elongated system and $\sigma/\zeta$ is much smaller than $P_b$ and $P_f$ in experimental conditions, we simply use $c_\sigma=0$ for the discussions about the overlap integral. In the PSL model of the monopole mode, we calculate for both $c_\sigma=0$ (TF limit) and $c_\sigma=1$ (cylindrical approximation).

Using Eq.~(\ref{eq:Nb}-\ref{eq:pressure}) and evaluating $\sigma$ with Eq. \eqref{eq:sigma} at the interface, we can fix $R_b$, $R_f$ and $\zeta$ and get the density distributions at the PSL.

\section{collective mode of BEC}
The wave equation of BEC is 
\begin{equation}\label{eq:PS_BEC-wave}
\omega^2 \delta n_b + \frac{g_{bb}}{m}\nabla \cdot \left(n_b\nabla  \delta n_b\right) =0,
\end{equation}
where $n_b(r)$ is BEC density at equilibrium and $\delta n_b$ is the density fluctuation. 
While the solution for a cylindrical system is given in Eq.~\eqref{eq:2F1_cyl},
the solutions in a spherically symmetric trap is
\begin{equation}
\delta n_b(r)\propto Y_l^m r^l F(\alpha^+,\alpha^-,l+3/2,(r/R_b)^2),
\end{equation} with $F$ the hypergeometric function $_2F_1$, $Y_l^m$ the spherical harmonic, $2\alpha^{\pm}=z\pm[z^2+2(\omega^2/\omega_b^2-l)]^{1/2}$, $z=l+3/2$ and $l \in \mathbb{N}$. Here we only need the case with $m=l=0$.

\section{Eigenvalue equation of the breathing mode}\label{app:osci_dp}
\subsection{cylindrical condition}
The pressure balance at the interface with a displacement of $\delta \zeta$ is
\begin{equation}\label{eq:pressure_app0}
\begin{split}
\frac{\partial P_b}{\partial\mu_b} \delta \mu_b  -\delta \Pi +\left(\frac{\partial P_b}{\partial\zeta}-\frac{\partial P_f}{\partial\zeta}\right)\delta \zeta\\
=-\frac{\sigma}{\zeta^2}\delta \zeta	+\frac{\partial \sigma}{\partial\mu_b}\frac{\delta\mu_b}{\zeta}+\frac{\partial \sigma}{\partial\zeta}\frac{\delta\zeta}{\zeta}.
\end{split}
\end{equation}
From Eq.~\eqref{eq:BC_BEC} we find the relation between $\delta\mu_b=g_{bb}\delta n_b$ and $\delta \zeta$ as 
\begin{equation}
\delta n_b=\frac{F}{\partial_r F}\frac{m_b \omega^2}{g_{bb}}\delta\zeta.
\end{equation}
Together with other relevant replacing rules
\begin{equation}\label{eq:boundary_eqs}
\begin{split}
\frac{\partial P_b}{\partial\mu_b}\delta\mu_b&=g_{bb}n_b\delta n_b\\ 
\frac{\partial P_b}{\partial\zeta}&=P_b\frac{-4\zeta}{R_b^2-\zeta^2}\\
\frac{\partial P_f}{\partial\zeta}&=P_f\frac{-5\zeta}{R_f^2-\zeta^2}\\
\delta \Pi&=\frac{p_F^4}{(2\pi\hbar)^3}C_\Pi \delta \zeta,
\end{split} 
\end{equation}
Eq.~\eqref{eq:pressure_app0} becomes Eq.~\eqref{eq:BC_PSL_cyl}. And we can further simplify  Eq.~\eqref{eq:BC_PSL_cyl} to have
\begin{equation}\label{eq:BC_PSL_cyl_numerical}
\frac{F_x}{F}=\frac{R_b m_b \omega^2 \left(n_b-{C_\sigma}\frac{ \sigma}{\mu_b \zeta}\right)}{-\frac{\sigma}{\zeta^2}- C_\sigma\frac{m_b\omega_b^2 \sigma}{2 \mu_b} +\frac{p_f^4 C_{\Pi}}{(2\pi\hbar)^3}-\partial_r(P_b-P_f)},
\end{equation}
where we define formally $\partial F(\alpha^+,\alpha^-,1,x^2)/\partial x=F_x$. If we apply the adiabatic fermionic reservoir approximation ($\delta \Pi=0$) and ignore the surface tension terms, Eq.~\eqref{eq:BC_PSL_cyl_numerical} becomes extremely simple as $\omega^2 F=x F_x$. In the limit that $x$ approaches zero, we get approximately $\omega/\omega_b\approx c_J/(\sqrt{2}x)$, where the constant $c_J\approx2.4048$ is the first root of the Bessel function $J_0$ and $x \propto (N_b/N_f^{25/24})^{1/3}$ for our harmonically trapped mixture.

\subsection{Spherical condition}
For fermions the deformation of Fermi surface is $f=f_0+\delta(|\vec{p}|-p_f)u(r,\chi)e^{-i\omega t}$, where $\chi=\cos\theta$ with $\theta$ the angel between $\vec{p}$ and $\vec{r}$ and $p_f(r)=\sqrt{2 m_f\mu_f(r)}$ is the local Fermi momentum. In the case of spherical system, we get
\begin{equation}
-i\omega u +\omega_f\rho\chi\partial_r u +\omega_f(1-\chi^2)g(r)\partial_\chi u=0,
\end{equation} where $\rho=\sqrt{R_f^2-r^2}$ and $g(r)=\rho/r-r/\rho$.
The corresponding solution has the form 
\begin{equation}
u(r,\chi)=\mathcal{F}[r^2(R_f^2-r^2)(1-\chi^2)]e^{-i\omega \tau/2},
\end{equation}
where $\mathcal{F}$ is an arbitrary function and
\begin{equation}
\tau(\zeta,\chi)=\frac{\phi_0 -\arctan[2\chi/g(\zeta)]}{\omega_f},
\end{equation}
where the constant $\phi_0$ is fixed by the actual physics of the system.

As $\tau$ characterizes the phase of the oscillating system, we recognize it as the time for a single fermion to depart and then return to the interface in case $\chi\in[0,1]$, and have $\tau(\zeta,\chi)=-\tau(\zeta,-\chi)$. 

We also have to apply the non-penetration boundary condition for fermions at $r_e=\zeta$ that 
\begin{equation}
\left[u(\zeta,\chi)-u(\zeta,-\chi)\right]e^{-i\omega t}=2 m_f \chi \dot{\zeta},
\end{equation} where $\dot{\zeta}=\partial_t \zeta=\partial_t(\delta\zeta e^{-i\omega t})$ is the velocity of the phase boundary. Finally, the fermion perturbation is solved to be
\begin{equation}\label{eq:u_sph}
u(\zeta,\chi)=2m_f\chi(-i\omega\delta\zeta)(1-e^{i\omega\tau})^{-1}.
\end{equation}

The equilibrium condition of pressure at the phase boundary involves two variables $\mu_b$ and $\zeta$, and it is equivalent to consider the pressure balance in the lab frame or in a moving frame with the surface speed $\dot{\zeta}$. We assume the BEC evolves in a quasi-static process and the equation of state is still valid, which gives $P_b=g_{bb}n_b^2/2$ and $n_b$ depends on the oscillation of density via $\delta\mu_b=g_{bb}\delta n_b$ and the variation of the surface position $\delta \zeta$.  On the other hand when $|\dot{\zeta}|\ll v_F$, the Fermi pressure fluctuation at the boundary is given by the radial momentum flux element 
\begin{equation}
\delta\Pi_{rr}(\zeta)=\frac{1}{m_f(2\pi\hbar)^3}\int d^3 p (\chi p)^2 (f-f_0),
\end{equation}
where $\delta f=f-f_0$ is the perturbation of the fermion distribution function.
Consider only the monopole mode for spherically symmetric system and drop out $e^{-i\omega t}$ on both side, we have
\begin{equation}
\begin{split}
\frac{\partial P_b}{\partial\mu_b} \delta \mu_b  -\delta \Pi +\left(\frac{\partial P_b}{\partial\zeta}-\frac{\partial P_f}{\partial\zeta}\right)\delta \zeta\\
=-\frac{2\sigma}{\zeta^2}\delta \zeta	+2\frac{\partial \sigma}{\partial\mu_b}\frac{\delta\mu_b}{\zeta}+2\frac{\partial \sigma}{\partial\zeta}\frac{\delta\zeta}{\zeta}.
\end{split}
\end{equation}
We use Eq.~\eqref{eq:boundary_eqs} with ${C_{\Pi}(\omega,\omega_f)}={4\pi\omega \int_{0}^{1} d\chi \chi^3
\cot\left({\omega \tau}/{2 }\right)}$. Finally the equation for $\omega$ in the spherical case is
\begin{equation}\label{eq:BC_fermion_sphere}
\frac{\partial_r \delta n_b}{\delta n_b}=\frac{\omega^2 m_b\left(n_b-C_\sigma\frac{2\sigma}{\mu_b\zeta}\right)}{-\frac{2\sigma}{\zeta^2}-C_\sigma\frac{m_b \omega_b^2\sigma}{\mu_b}+\frac{p_f^4 }{(2\pi\hbar)^3}C_{\Pi}-\partial_r(P_b-P_f)}.
\end{equation}

\bibliographystyle{apsrev4-1}

\end{document}